# Large deformation and post-failure simulations of segmental retaining walls using mesh-free method (SPH)

Simulations de grandes déformations et post-rupture des murs de soutènement segmentaires utilisant la méthode des mailles-libres (SPH)

Bui H.H., Kodikara J.A, Pathegama R., Bouazza A., Haque A.
*Department of Civil Engineering, Monash University, Melbourne, Vic 3800, Australia.*

ABSTRACT: Numerical methods are extremely useful in gaining insights into the behaviour of reinforced soil retaining walls. However, traditional numerical approaches such as limit equilibrium or finite element methods are unable to simulate large deformation and post-failure behaviour of soils and retaining wall blocks in the reinforced soil retaining walls system. To overcome this limitation, a novel numerical approach is developed aiming to predict accurately the large deformation and post-failure behaviour of soil and segmental wall blocks. Herein, soil is modelled using an elasto-plastic constitutive model, while segmental wall blocks are assumed rigid with full degrees of freedom. A soft contact model is proposed to simulate the interaction between soil-block and block-block. A two dimensional experiment of reinforced soil retaining walls collapse was conducted to verify the numerical results. It is shown that the proposed method can simulate satisfactory post-failure behaviour of segmental wall blocks in reinforced soil retaining wall systems. The comparison showed that the proposed method can provide satisfactory agreement with experiments.

RÉSUMÉ : Les méthodes numériques sont extrêmement utiles pour obtenir un aperçu du comportement des murs de soutènement en sol renforcé. Cependant, les approches numériques traditionnelles tels que l'équilibre limite ou méthodes d'éléments finis sont incapables de simuler les déformations importantes et le comportement post-rupture des sols et les blocs de béton des murs de soutènement segmentaires. Pour contourner cette limitation, une nouvelle approche numérique est présentée dans cet article. Le sol est modélisé à l'aide d'un modèle élasto-plastique, tandis que les blocs segmentaires muraux sont supposés être rigides avec une degré de liberté total. Un modèle de contact souple a été développé pour modéliser l'interaction entre le sol-bloc et bloc-bloc. Un modèle expérimentale en deux dimensions d'un effondrement d'un mur de soutènement renforcé a été réalisé pour vérifier les résultats numériques. Il est montré que la méthode de simulation proposée permet de simuler le comportement post-rupture des blocs de mur segmentaires des murs de soutènement renforcé. La comparaison a montré que la méthode proposée peut fournir un accord satisfaisant avec les résultats expérimentaux.

## 1 INTRODUCTION

In recent years, segmental retaining walls (SRWs) have received great attentions for their low material cost, short construction period, ease of construction, and aesthetic appearance. They have been used as an effective method to stabilize cuts and fills adjacent to highways, and embankments, amongst many other applications. Because of the flexible structural materials used (no mortar, or concrete footing), SRWs can tolerate minor ground movement and settlement without causing damage or cracks. In addition, dry stacked SRW construction allows free draining of water through the wall face, thereby reducing hydrostatic pressure build-up behind the wall.

Thus far, several analytical and numerical approaches have been developed to assist SWR design. Among these techniques, the finite element method (FEM) has been frequently applied to investigate stability and settlement of segmental retaining wall systems. FEM has also been used to simulate seismic load-induced large deformation of SRW systems. However, because of the mesh-based nature, FEM suffers from mesh entangling when dealing with large deformation problems, even when the updated Lagrangian method is adopted. Re-meshing may help to resolve this problem but the procedure is quite complicated. It is also worth mentioning that the free rotation motion of retaining wall blocks in SWR systems could not be modeled by FEMs. As an alternative for such computational complications, it is attractive to develop mesh-free methods. So far, the most popular mesh-free method applied in geotechnical engineering is the discrete element method (DEM) which tracks the motion of a large number of particles, with inter-particle contacts modelled by spring and dashpot systems (Cundall & Strack, 1979). The main advantages of this approach are that it can handle large deformation and failure problems; and the concept is relatively simple and easy to implement in a computer code. Thus, DEM could be considered an ideal method to simulate the full degrees-of-freedom motion of the retaining wall blocks in SRW systems. However, to model soil behaviour, DEM suffers from low accuracy because suitable parameters for the contact model are difficult to determine. The discontinuous deformation analysis (DDA) method proposed by Shi et al. (1998) has also been applied to geotechnical applications, but is mainly used for rock engineering, etc. Other continuum based mesh-free methods such as the mesh-free Galerkin element method (EFG), material point method (MPM), particle in cell method (PIC), etc., could be also applied to simulate large deformation of soil. However, these methods are quite time consuming and complicated to implement into a computer code as they consist of both interpolation points and the background mesh. On the other hand, the smoothed particle hydrodynamics (SPH) method, originally proposed by Gingold & Monaghan (1977), has been recently developed for solving large deformation and post-failure behaviour of geomaterials (Bui et al. 2007-2012; Pastor et al. 2009, Blanc et al. 2012) and represents a powerful way to understand and quantify the failure mechanisms of soil in such challenging problems. In this paper, taking into consideration this unique advantage, SPH is further extended to simulate large deformation and post-failure of SRW systems.

## 2 SIMULATION APPROACHES

### 2.1 Soil modelling in SPH framework

In the SPH method, motion of a continuum is modeled using a set of moving particles (interpolation points); each assigned a constant mass and "carries" field variables at the corresponding location. The continuous fields and their spatial derivatives are taken to be interpolated from the surrounding particles by a weighted summation, in which the weights diminish with distance according to an assumed kernel function. Details of the interpolation procedure and its application to soil can be found in Bui et al. (2008). The motion of a continuum can be described through the following equation,

$$\rho \ddot{\mathbf{u}} = \nabla \boldsymbol{\sigma} + \rho \mathbf{g} + \mathbf{f}_{ext} \qquad (1)$$

where $\mathbf{u}$ is the displacement; a raised dot denotes the time derivative; $\rho$ is the density; $\boldsymbol{\sigma}$ is the total stress tensor, where negative is assumed for compression; $\mathbf{g}$ is the acceleration due to gravity; and $\mathbf{f}_{ext}$ is the additional external force(s). The total stress tensor of soil is normally composed of the effective stress ($\boldsymbol{\sigma}'$) and the pore-water pressure ($p_w$), and follows Terzaghi's concept of effective stress. Because the pore-water pressure is not considered, the total stress tensor and the effective stress are identical throughout this paper and can be computed using a constitutive model.

In the SPH framework, Equation (1) is often discretized using the following form,

$$\ddot{u}_a^\alpha = \sum_{b=1}^{N} m_b \left( \frac{\sigma_a^{\alpha\beta}}{\rho_a^2} + \frac{\sigma_b^{\alpha\beta}}{\rho_b^2} + C_{ab}^{\alpha\beta} \right) \nabla_a^\beta W_{ab} + g_a^\alpha + f_{ext \to a}^\alpha \qquad (2)$$

where $\alpha$ and $\beta$ denote Cartesian components $x$, $y$, $z$ with the Einstein convention applied to repeated indices; $a$ indicates the particle under consideration; $\rho_a$ and $\rho_b$ are the densities of particles $a$ and $b$ respectively; $N$ is the number of "neighbouring particles", i.e. those in the support domain of particle $a$; $m_b$ is the mass of particle $b$; $C$ is the stabilization term employed to remove the stress fluctuation and tensile instability (Bui et al., 2011); $W$ is the kernel function which is taken to be the cubic Spline function (Monaghan & Lattanzio 2005); and $f_{ext \to a}$ is the unit external force acting on particle $a$.

The stress tensor of soil particles in Equation (2) can be computed using any soil constitutive model developed in the literature. For the purpose of soil-structure interaction, the Drucker-Prager model has been chosen with non-associated flow rule, which was implemented in the SPH framework by Bui et al. (2008) and shown to be a useful soil model for simulating large deformation and post-failure behaviour of aluminum rods used in the current paper as model ground. The stress-strain relation of this soil model is given by,

$$\dot{\boldsymbol{\sigma}} = \mathbf{D}^e : (\dot{\boldsymbol{\varepsilon}} - \dot{\boldsymbol{\varepsilon}}^p) \qquad (3)$$

where $\mathbf{D}^e$ is the elastic constitutive tensor; $\dot{\boldsymbol{\varepsilon}}$ is the strain rate tensor; and $\dot{\boldsymbol{\varepsilon}}^p$ is its plastic component. An additive decomposition of the strain rate tensor has been assumed into elastic and plastic components. The plastic component can be calculated using the plastic flow rule,

$$\dot{\boldsymbol{\varepsilon}}^p = \dot{\lambda} \frac{\partial g_p}{\partial \boldsymbol{\sigma}} \qquad (4)$$

where $\dot{\lambda}$ is the rate of change of plastic multiplier, and $g_p$ is the plastic potential function. The plastic deformation occurs only if the stress state reaches the yield surface. Accordingly, plastic deformation will occur only if the following yield criterion is satisfied,

$$f = \alpha_\phi I_1 + \sqrt{J_2} - k_c = 0 \qquad (5)$$

where $I_1$ and $J_2$ are the first and second invariants of the stress tensor; and $\alpha_\phi$ and $k_c$ are Drucker-Prager constants that are calculated from the Coulomb material constants $c$ (cohesion) and $\phi$ (internal friction). In plane strain, the Drucker-Prager constants are computed by,

$$\alpha_\phi = \frac{\tan\phi}{\sqrt{9 + 12\tan^2\phi}} \quad \text{and} \quad k_c = \frac{3c}{\sqrt{9 + 12\tan^2\phi}} \qquad (6)$$

The non-associated plastic flow rule specifies the plastic potential function by,

$$g_p = \alpha_\psi I_1 + \sqrt{J_2} - \text{constant} \qquad (7)$$

where $\alpha_\psi$ is a dilatancy factor that can be related to the dilatancy angle $\psi$ in a fashion similar to that between $\alpha_\phi$ and friction angle $\phi$. Substituting Equation (7) into Equation (4) in association with the consistency condition, that is the stress state must be always located on the yield surface $f$ during the plastic loading, the stress-strain relation of the current soil model at particle $a$ can be written as,

$$\frac{d\sigma_a^{\alpha\beta}}{dt} = 2G_a \dot{e}_a^{\alpha\beta} + K_a \dot{\varepsilon}_a^{\gamma\gamma} \delta^{\alpha\beta} - \dot{\lambda}_a \left[ 3K_a \alpha_{\psi a} \delta^{\alpha\beta} + (G_a/\sqrt{J_{2a}}) s_a^{\alpha\beta} \right] \qquad (8)$$

where $e^{\alpha\beta}$ is the deviatoric strain-rate tensor; $s^{\alpha\beta}$ is the deviatoric shear stress tensor; and $\dot{\lambda}$ is the rate of change of plastic multiplier of particle $a$, which in SPH is specified by

$$\dot{\lambda}_a = \frac{3\alpha_{\phi a} K_a \dot{\varepsilon}_a^{\gamma\gamma} + (G_a/\sqrt{J_{2a}}) s_a^{\alpha\beta} \dot{\varepsilon}_a^{\alpha\beta}}{9\alpha_{\phi a} K_a \alpha_{\psi a} + G_a} \qquad (9)$$

where the strain-rate tensor is computed by

$$\dot{\varepsilon}_a^{\alpha\beta} = \frac{1}{2} \left( \nabla^\beta \dot{u}^\alpha + \nabla^\alpha \dot{u}^\beta \right)_a \qquad (10)$$

When considering a large deformation problem, a stress rate that is invariant with respect to rigid-body rotation must be employed for the constitutive relations. In the current study, the Jaumann stress rate is adopted:

$$\dot{\sigma}_a^{\alpha\beta} = \dot{\sigma}_a^{\alpha\beta} - \sigma_a^{\alpha\gamma} \dot{\omega}_a^{\beta\gamma} - \sigma_a^{\gamma\beta} \dot{\omega}_a^{\alpha\gamma} \qquad (11)$$

where $\dot{\omega}$ is the spin-rate tensor computed by

$$\dot{\omega}_a^{\alpha\beta} = \frac{1}{2} \left( \nabla^\beta \dot{u}^\alpha - \nabla^\alpha \dot{u}^\beta \right)_a \qquad (12)$$

As a result, the final form of the stress-strain relationship for the current soil model is modified to

$$\frac{d\sigma_a^{\alpha\beta}}{dt} = \sigma_a^{\alpha\gamma} \dot{\omega}_a^{\beta\gamma} + \sigma_a^{\gamma\beta} \dot{\omega}_a^{\alpha\gamma} + 2G_a \dot{e}_a^{\alpha\beta} + K_a \dot{\varepsilon}_a^{\gamma\gamma} \delta_a^{\alpha\beta} - \dot{\lambda}_a \left[ 3K_a \alpha_{\psi a} \delta^{\alpha\beta} + (G/\sqrt{J_2})_a s_a^{\alpha\beta} \right] \qquad (13)$$

Equations (2) and (13) are finally integrated using Leapfrog algorithm to describe the motion of soil medium. Validation of this soil model with SPH has been extensively documented in the literature (Bui et al. 2008-2012), and reader can refer to these references for further details on the validation process.

2.2 *Rigid body motion of retaining wall blocks*

The segmental retaining wall simulated in this paper is comprised of individual rectangular blocks; each is assumed as a rigid body and has complete degrees-of-freedom. The motion of the block can be determined by specifying the motion of the central mass and the rotation about its mass central. The equation of motion of the central mass is given as follows,

$$M\frac{dV}{dt} = F \qquad (14)$$

where $M$ is the central mass, $V$ is the velocity vector of the central mass, $F$ is total force vector acting on the body. The equation of rotation about the central mass is,

$$I\frac{d\Omega}{dt} = T \qquad (15)$$

where $I$ is the inertial moment, $\Omega$ is the angular velocity which is perpendicular to the plane of the motion, and $T$ is the total torque about the central mass.

In the computation, the rectangular block is represented by the set of boundary particles that are equi-spaced around the boundary. Denoting the force vector acting on each boundary particle $i$ located on the moving block is $f_i$, Equations (14) and (15) can be rewritten, respectively, as follows,

$$M\frac{dV}{dt} = \sum_i f_i \qquad (16)$$

$$I\frac{d\Omega}{dt} = \sum_k (r_i - R) \times f_i \qquad (17)$$

where $r_i$ and $R$ are vector coordinates of boundary particle and central mass, respectively. The rigid body boundary particles move as a part of the rigid body, thus the change on position of boundary particle $i$ is given by,

$$\frac{dr_i}{dt} = V + \Omega \times (r_i - R) \qquad (18)$$

The force $f_i$ acting on a boundary particle on the rigid body is due to the surrounding soil particles or boundary particles that belong to different rigid bodies. This force can be calculated using a suitable contact model which will be described in the next section.

2.3 *Contact force model*

In this paper, a soft contact model based on a concept of the spring and dash-pot system is proposed to model the interaction between soil and retaining wall blocks and between blocks. Accordingly, the radial force acting between two particles can be calculated using the following equation,

$$f_{a\rightarrow i}^n = \begin{cases} -K_{ai}\delta_n^{3/2} - c_n v_{ak}^n & (h_a + h_i) \le 2d_{ai} \\ 0 & (h_a + h_i) > 2d_{ai} \end{cases} \qquad (19)$$

where $K$ is the radial stiffness; $\delta_n$ is the allowable overlapping distance between two particles; $c_n$ is the radial damping coefficient; $v^n$ is the relative radial velocity vector between particle $a$ and particle $i$; $h_a$ and $h_i$ are the initial distance (so-called smoothing length in SPH) between soil particles and between boundary particles, respectively; and $d_{ai}$ is the distance between two particles. The stiffness, overlapping distance and damping coefficient can be calculated using the following relationships,

$$K_{ai} = E_{eff}\sqrt{h_{eff}}/3 \qquad (20)$$

$$\delta_n = d_{ai} - (h_a + h_i)/2 \qquad (21)$$

$$c_n = 2\sqrt{m_{ai}K_{ai}} \qquad (22)$$

where $E_{eq}$ and $h_{eq}$ are equivalent Young's modulus and smoothing length, respectively. The contact force in the shear direction which is perpendicular to the radial direction can be calculated in the same manner,

$$f_{a\rightarrow i}^s = \begin{cases} -k_{ai}\delta_s - c_s v_{ai}^s & (h_a + h_i) \le 2d_{ai} \\ 0 & (h_a + h_i) > 2d_{ai} \end{cases} \qquad (23)$$

where $k_{ai}$ is the shear stiffness; $\delta_s$ is the relative displacement between the two particles in the shear direction; $c_s$ is the shear damping coefficient; $v^s$ is the relative shear velocity vector between particle $a$ and particle $i$. These unknown variables can be calculated using the following relationships,

$$k_{ai} = 4G_{eq}\sqrt{h_{eq}\delta_n} \qquad (24)$$

$$\delta_s = \oint v_{ai}^s dt \qquad (25)$$

$$c_n = 2\sqrt{m_{ai}k_{ai}} \qquad (26)$$

where $G_{eq}$ is the equivalent shear modulus. The current shear force must satisfy Coulomb's friction law which implies that the maximum shear force must not exceed the maximum resisting force,

$$f_{a\rightarrow i}^s \le \mu \frac{\delta_s}{|\delta_s|}\left|f_{a\rightarrow i}^n\right| \qquad (27)$$

Finally, these forces are converted to the conventional coordinate system and added to Equations (2), (16) and (17) to progress the motion of soil and rigid bodies.

3 OUTLINE OF MODEL TEST

Two-dimensional experiments of retaining wall collapse were conducted to validate the SPH numerical results. Figure 1 shows a schematic diagram of the two-dimensional experimental setup and Figure 2 shows the initial setup condition of the model ground and retaining wall blocks in the experiment. The size of the model ground is 15cm in height and 50cm in width at the base. Aluminum rods of 5cm in length, having diameters of 1.5 to 3mm and mixed with the ratio 3:2 in weights, are used as the model ground. The total unit weight of the model ground is 23kN/m³. The retaining block is 3.2cm in width, 2.5cm in height, and 5cm in length, which is manufactured from aluminum (Young's modulus of 69GPa and unit weight of 26.5kN/m³). In the experiment, the segmental retaining wall system was constructed by successively placing one block on the top of the other with an overlapping of 1.9cm, followed by filling the model ground at each layer. To visualize the failure pattern of the model ground, square grids (2.5×2.5cm) were drawn on the specimen. The experiments were initiated by quickly removing the stopper stand and digital photos were taken to record the failure process as well as the final configuration of the retaining wall system after collapse.

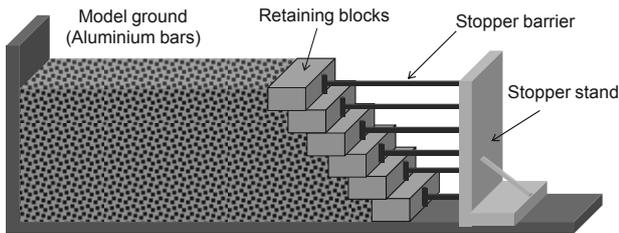

Figure 1. Schematic diagram of 2D experimental model.

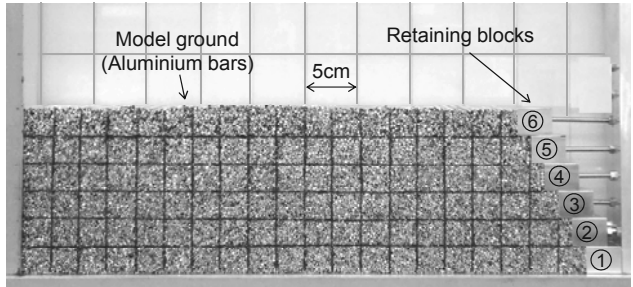

Figure 2. Initial setting condition of the model test.

In addition, tests were conducted to measure static friction coefficients. It was found that the static friction ($\mu$) between retaining wall blocks is ~ 0.62, between wall block and the bottom wall boundary is ~ 0.60, and between the retaining wall block and model ground is ~ 0.56. The experiment series were conducted starting from one block and then gradually increasing the number of blocks in the retaining wall system until the retaining wall collapsed. It was observed that the retaining wall system collapsed when reaching 6 blocks. Accordingly, a numerical model consisting of 6 retaining wall blocks was conducted for the benchmark study.

## 4 RESULT AND DISCUSSION

The model test shown in Figure 2 was simulated using 11,304 SPH particles arranged in a rectangular lattice with an initial separation of 0.25cm. Rigid blocks were created by placing boundary particles uniformly around the boundary at a constant distance. In order to simulate the smooth surface, half of particle spacing was chosen for the rigid body boundary particles. Model ground parameters including elastic modulus $E = 1.5$MPa, Poisson's ratio $\nu = 0.3$, friction angle $\phi = 19.8°$, dilatant angle $\psi = 0°$, and cohesion $c = 0$kPa were taken similar to those used in Bui et al. (2008). The unit weight of the ground model is $\gamma_s = 23$kN/m$^3$. The friction coefficients between the rigid blocks, between the block and the bases of the wall boundary, and between the block and the ground model were taken to be similar to those measured in the experiment as mentioned in Section 3. The boundary conditions for the model ground are restrained with a free-slip boundary at the lateral boundaries and fixed in both directions at the base.

Figure 3 shows the comparison between the experiment and the computation. It can be seen that, because the complete degrees-of-freedom of the rigid body was taken into consideration, the computed result could predict fairly well the behaviour of all rigid blocks observed in the experiment after the SWR system collapsed. The final run-out distance of Block No.1 in the simulation is 67.5cm from the left-most solid boundary. This result is in very good agreement with that observed in experiment (~68cm). It is suggested that the proposed soft contact model could be applied to simulate the soil-structure interaction in the SRW system. However, further refinement of the contact model should be considered to provide more accurate prediction of the retaining wall blocks and the model ground as the simulation results showed a slight over-prediction of the failure zone observed in the experiment.

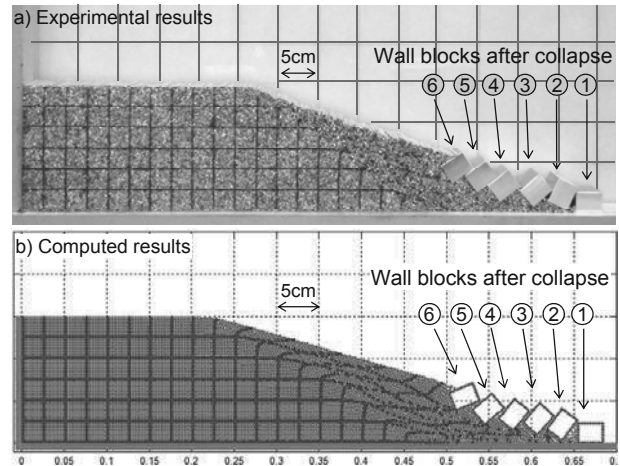

Figure 3. Comparison between experiment and SPH simulation.

## 5 CONCLUSION AND FUTURE WORKS

This paper presented a novel numerical approach for simulation of large deformation and post-failure of segmental retaining wall. It was shown that the proposed method provides good agreement with the experimental results. The most significant advantage of the new method is that the complete degrees-of-freedom of the retaining wall blocks, which could not be simulated using traditional numerical approaches, can now be simulated in the proposed numerical framework. Large deformation and post-failure behaviour of geomaterials can also be readily simulated. To broaden the application of the proposed numerical approach, further implementations such as coupling with geo-grid reinforcement, modelling seismic earthquake loading, and bonding between blocks should be considered in the future. Full extension to three-dimension code would yield significant benefits to gain further insights into the mechanisms of SWR.